# Solving it correctly: Prevalence and Persistence of Gender Gap in Basic Mathematics in rural India

**Authors**
Upasak Das and Karan Singhal

**Affiliation**

(i) Upasak Das (corresponding author)

Presidential Fellow of Economics of Poverty Reduction
Global Development Institute
University of Manchester, United Kingdom

Research affiliate
Centre of for Social Norms and Behavioral Dynamics
University of Pennsylvania

*Email:* upasak.das@manchester.ac.uk

*Full address:*
Room 1.008
Global Development Institute
Arthur Lewis Building, Oxford Road,
University of Manchester,
Manchester- M13 9PL, United Kingdom

(ii) Karan Singhal

Project Officer
Indian Institute of Management Ahmedabad, India
Email: karansinghal1993@gmail.com

**Acknowledgements**

We thank Ambrish Dongre, Anuradha De, Karthikeya Naraparaju, Udayan Rathore, Amartya Paul, Nisha Vernekar and Amarendra Kumar along with the seminar participants at Indian Statistical Institute Delhi, South Asian University, Delhi School of Economics, Jadavpur University Kolkata, IMERT Pune and National University of Singapore, for helpful comments and suggestions. We also thankful to the ASER Centre for sharing the data with us.

**Conflict of Interest Statement**

We would like to confirm that there are no known conflicts of interest associated with this work and there has been no significant financial support that could have influenced its outcome.

**Abstract**

*Mathematical ability is among the most important determinants of prospering in the labour market. Using multiple representative datasets with learning outcomes of over 2 million children from rural India in the age group 8 to 16 years, the paper examines the prevalence of gender gap in performance in mathematics and its temporal variation from 2010 to 2018. Our findings from the regressions show significant gender gap in mathematics, which is not observable for reading skills. This difference in mathematics scores remains prevalent across households of different socio-economic and demographic groups. This gap is found to be persistent over time and it appears to increase as the children get older. We also find significant inter-state variation with the north Indian states lagging behind considerably and the south Indian states showing a 'reverse gender gap'. As an explanation to this, we observe evidence of a robust association between pre-existing gender norms at the household and district level with higher gender gap. The findings, in light of other available evidence on the consequences of such gaps, call for the need to understand these gender specific differences more granularly and periodically to inform gender- specific interventions.*

## I. INTRODUCTION

The importance of learning abilities in mathematics remains crucial especially on future prospects in the labour market. Extant literature has documented gender gaps in learning, particularly mathematics could be a key explanatory factor in differences in pursuance of STEM discipline among boys and girls (Breda and Napp 2019), and by extension negatively affect women in future labour market outcomes (Altonji and Blank, 1999; Hanushek and Woessemann, 2008; Dossi, Figlio, Giuliano, and Sapienza 2021). Therefore, studying the performance in basic mathematical learning becomes important with respect to policies focusing on skill development and abilities that are conducive to labour market opportunities. Thus, unsurprisingly, gender differences in performance in mathematics have received significant attention in the past two decades, especially in middle and higher income countries, however evidence for developing countries still remains scarce (Fryer and Levitt 2010; Bharadwaj et al. 2012; Bharadwaj et al. 2016; Lippman and Senik 2018).

In this paper, we provide evidence on the prevalence and persistence of gender gaps in mathematics among adolescent children from rural India. In particular, we assess the following: is there an evident gender difference in the levels of basic mathematical abilities? Does this gap remain persistent across children from households that vary in terms of religion, social identity or economic factors among others? Does this gap remain persistent across time? Is there a convergence or divergence in the gender gap when considering children from younger to older age cohorts? How does the state-wise variation in gender differences in mathematical ability appear and how has it changed over time? And finally, how do gender norms and attitude that emanate at the household and society are linked to performance of the females.

This overview on gender differences in mathematics learning is highly relevant to the Indian context given very limited evidence on its size and persistence. There is currently sparse evidence on this and related dimensions. For example, Muralidharan and Seth (2016) examine the role of teacher gender in reducing gender gaps in learning outcomes in the state of Andhra Pradesh and find that girls are likely to score at par with boys in mathematics at the end of first grade but perform significantly worse by the end of grade 5. Rakshit and Sahoo (2020) use data from the states of Andhra Pradesh and Telangana and find biased teachers to negatively impact attitudes of girls towards maths relative to boys. Singh and Krutikova (2017) find no gender gaps in learning among boys and girls at the pre-primary and primary levels but an increasing gap in the post primary stage across four countries, including India based on the same states of Andhra Pradesh and Telangana. However, none of the studies have focused solely on differences in basic mathematical learning or documented its persistence and spatial variation over the years, which is where this paper gains significance and remains its main contribution.

The case of India is pertinent for several reasons. Firstly, the situation of primary and secondary education in the country remains adverse. Despite achieving close to universal primary enrolment, reports document that 29 and 43 percent of children drop out before completing five years of primary school, and finishing upper primary school respectively, which is more common among females. The same report finds a teacher shortage of 689,000 teachers in primary schools with only 47 percent of the schools not having functional toilets for girls and 26 percent not having access to drinking water. Secondly, data suggests that gender inequality in India is found to be even higher than some of the other countries with lower human development indicators like Rwanda, Ethiopia or Burundi. For example, Gender Inequality Index (GII) in Rwanda is found to be 0.383 in 2015, whereas for India, it is as high as 0.530 and this figure is likely to be higher in rural India.[2] In one of the recent studies, it is found that India has slipped many places in terms of Global Gender Gap Index brought out by the World Economic Forum[3] and there are evidence to suggest that gender gap in mathematics is higher countries with greater gender divide (Guiso, Monte, Sapienza, and Zingales 2008).

The paper draws insights from the Annual Status of Education Report (ASER) and supplements it with two additional sources – the Indian Human Development Survey (IHDS) conducted in 2011-12 and the fourth round of National Family Health Survey (NFHS-4). All these three surveys offer nationally representative information with the first two providing data on performance of basic abilities in mathematics and reading among children.[1] National Achievement Survey (NAS) is the only other source which provides national level learning estimates but lacks available household data and not as reliable for comparison across states because of 'unrealistically high' averages (Johnson and Parrado, 2021).

First, we utilize the learning outcomes data for rural India for children from 8 years to 16 years of age from the annual ASER survey starting from 2010 till 2018. We further complement this by utilising household data from IHDS that provides information on socioeconomic and demographic characteristics which are not available in the ASER survey. This allows us to examine if the gender gap remains prevalent in households that differ by these characteristics. Next, we use the ASER dataset from 2010 to 2018 to gauge whether the gap converges or diverges over time giving an indication of improvement (if any) over the eight years. Additionally, we also assess the state wise variation in the gender gap overall as well as across years using the ASER dataset. Finally, we use the first round of the IHDS survey and the NFHS-4 dataset to assess the implications of the prevailing gender gap in households with entrenched gender norms.

---

[1] IHDS also provides data on writing abilities.

We find a statistically significant and robust gender gap (in favour of males) that is evident in basic mathematical abilities, which is not found for reading abilities. This gap is found to persist over time and also prevalent among households of different socio-economic structure and demographic composition. Further, we find no evidence of convergence in the gap when we look at the difference across age - females at higher age are found to score consistently lesser than the male children of same age and this gap is higher than that for children of lower age.

Expectedly, we find large inter-state variation in this gender gap. South Indian states of Kerala, Tamil Nadu, Andhra Pradesh and Karnataka are observed to exhibit a reverse gender gap while the North Indian states that include Uttar Pradesh, Bihar and Madhya Pradesh among others are found to have a persistent and substantial gap. In addition, we find evidence of females from regressive gender norms within their household or society being more likely to systematically fare worse in mathematics, thus providing some explanation on the persistence of the gender gap in these north Indian states infamous for entrenched norms biased against women.

The contribution of our study lies in the fact that it gives an overview of gender gap in mathematical ability and documenting the prevalence and persistence of these gaps across age, time and geographical locations. While our work is unable provide causal channels for these differences, we also posit gendered norms and attitude as plausible mechanisms that can explain the prevailing gender gap and accordingly inform policy interventions in the area. Important to note here is, despite some glaring differences, any formal recognition of these gaps in policy also remains limited. The National Curriculum Framework (NCF) set up by the National Council of

Educational Research and Training (NCERT)[2] in India acknowledges the possibility of difference in the learning experience between males and females and higher mathematical phobia or anxiety among female children (NCERT, 2005). Yet, no discernible concrete steps to lessen this gap have been taken up either at the central or the state level. The recently passed New Education Policy in 2020, despite certain innovative steps to address issues pertaining to gender equity in education and the push towards Foundational Literacy and Numeracy (FLN) (meant to strengthen critical reading and mathematics among children up to grade 3) does not acknowledge this gender based difference. Our findings underscore persistence of the problem and put emphasis on states that should be forefront in ameliorating this issue. Our work should open discussion on the need to minimize the gap that can potentially lead to implementation of important and relevant policy instruments.

The paper has been structured as follows: Section 2 details the literature on the prevalence of gender differences in math scores and plausible reasons for gaps found in various contexts. Section 3 describes the data and variables used in the analysis and the empirical strategy. Section 4 describes the results in details, including the prevalence and persistence of differences in different socioeconomic contexts. Section 5 discusses the policy implications and conclusions coming out of the study.

## II. GENDER GAPS IN MATHEMATICS SCORES

Among the earliest work on the issue of gender gap in mathematics learning, Benbow and Stanley (1980) found the differences in the math component of the SAT examination. Recent studies continue to point to gender gaps in mathematics at various levels in developed as well as developing countries (Entwisle, Alexander, and Olson 1994; Bharadwaj et al. 2012; Bharadwaj et al. 2016; Ng'ang'a, Mureithi, and Wambugu, 2018). Lippman and Senik (2018) find these differences to be higher in western European countries than in the formally socialist eastern countries, attributing differences to the impact of social policies that pushed favourable gender norms in the eastern blocks. Guiso et al. (2008) using Programme for International Student Assessment (PISA) data from 40 countries find a positive correlation between the World Economic Forum's Gender Gap Index (covering aspects of economic participation, education attainment, and health indicators, among others) and gender gap in mathematics. Ellison and Swanson (2010), using data from American Math Competitions, find these differences to widen at the upper end of the distribution i.e. there are much fewer girls among top performers. Somewhat encouragingly, Hammond, Matulevich, Beegle, and Kumaraswamy (2020), based on data from Trends in International Mathematics and Science Study (TIMMS) and PISA find that the differences between math scores between boys and girls in the past four decades are closing, specifically in the top part of the distribution. Singh and Krutikova (2017) find no differences at the pre-primary and primary levels but growing differences in the post primary stage across favouring boys in India (Andhra Pradesh and Telangana), Ethiopia, and Peru, and favouring girls in Vietnam. Others too find these gaps to widen as children grow older across both developed and developing countries Fryer and Levitt, 2010; Bharadwaj et al. 2016; Muralidharan and Seth 2016, Smetackova, 2015).

---
[2] https://ncert.nic.in/nc-framework.php?ln= (accessed on 7th September 2021)

Some have argued that there is an innate difference in ability, brain development, and hormone levels in favour of boys which causes these gaps (Witelson 1976; Gur et al 1999; Davison and Susman 2001). However, more recent work attribute these differences in educational choices (in STEM or math related subjects at higher levels) and social factors (Antecol and Cobb-Clark, 2013; Zafar, 2013; Buser, Niederle, and Oosterbeek 2014; Kahn and Ginther, 2017; Smetackova 2015).

Cultural norms and gender based discriminatory practices at large can translate to differences in effort and performance in school (Bandhopadhyay and Subrahmanian 2008; Rodríguez-Planas and Nollenberger, 2018) and this may have a substantial bearing on differences in mathematics scores as well (Nollenberger, Rodríguez-Planas, and Sevilla 2016). Gender role stereotypes emanating from parents also explain some part of this difference in several outcomes (Eccles and Jacobs 1986; Eccles, Jacobs and Harold 1990; Parsons, Kaczala, and Meece 1982; Bhanot and Jovanovic 2005; Dhar, Jain, and Jayachandran 2018; Eble and Hu, 2018). In naturally occurring conversations, parents are found to be three times more likely to discuss science and related issues to boys in comparison to girls (Crowley, Callanan, Tenenbaum, and Allen 2001). Consequently, parents continue to believe that boys are better at math than girls and girls are made to believe that mathematics is not useful and is not a part of a girl's identity (Wilder and Powell, 1989; Jayachandran, 2015). Thus, not surprisingly, Devine, Fawcett, Szucs and Dowker (2012) find girls to have higher 'mathematical anxiety' (phobia or anxiety about one's ability to do math) than boys which affects them negatively in performance in mathematics. In this context, Dossi et al. (2021) find girls who grow up in boy-biased households score lower than other girls, and validate that such socialisation affects mathematical ability of girls.

Bias in the classroom from teachers and in education materials has also been noted in both developing and developed countries (Hammond et al. 2020). Bertrand (2011) finds that teachers may demand higher achievement from boys than girls while Carlana (2019) argues gender gap in mathematics among students in Italy increases for those who are assigned to teachers with stronger gender stereotypes. In India as well, studies have shown how gender stereotyping is deeply rooted in families and gender bias at home is a key element of the socialization process for girls (Mishra, Behara and Babu 2012). Lim and Meer (2017) find that female students are more likely to aspire to study STEM subjects and take advance math courses if they have a female teacher in middle/upper primary school. Similarly, in India, Rakshit and Sahoo (2020) show that biased teachers can negatively impact attitudes of girls towards maths.

### III. MATERIALS AND METHODS

Firstly, to examine the gender gap in learning outcomes, we make use of pooled household survey data from the ASER. The survey is district representative and conducted to study the schooling status as well as the basic levels of learning among children in rural India. It is conducted from the month of September to November every year in 550 out of the 720 districts in India and covers about 20 to 30 randomly sampled households from about 20 to 30 villages in each district. Accordingly, about 300,000 households are being surveyed in each round. For our analysis, we use data on over 2.3 million children surveyed from 2010 to 2018 in the age cohort 8 to 16 years.[3] In each of the surveyed households, all children in the age group 3 to 16 are surveyed and learning outcomes of children in the age group 5 to 16 are tested along with collecting information on their school enrollment among others.[4]

---

[3] The survey was annual earlier but now it is bi-annual has been conducted in the following years: 2010, 2011, 2012, 2013, 2014, 2016 and 2018.

[4] More information on ASER can be obtained from http://www.asercentre.org/ (accessed on September 15, 2019)

The survey also gathered information on basic arithmetic and reading proficiency levels using well tested rigorous tools.[5]. These tools are administered for all children across the districts and states and have been used extensively by other studies (Chakraborty and Jayaraman, 2018; Lahoti and Sahoo, 2020; Shah and Steinberg, 2019). The reading skill levels can be divided into four ordinal categories: identification of letters, identification of words, reading a grade 1 level text, and reading a grade 2 level text. The arithmetic skill levels comprise of four such categories: recognition of single-digit number, recognition of double-digit number, ability to do a two-digit subtraction, and ability to do a division (three by one form). For our analysis, we use standardized reading and mathematics scores that give a mean of 0 and a standard deviation of 1. Because a child of 8 years is expected to answer these levels correctly, we consider children in the age cohort 8 to 16 years for our analysis.[6]

Apart from these variables on learning outcomes, the survey also gathers household and village level information. We use those as independent variables in the regressions along with gender and age of the child. Household level characteristics include a number of indicators: whether the house is cemented or not and possession of television, mobile and toilets along with the household size. Following Banerjee et al. (2007), access to computers has been incorporated as a control. The village level factors include whether the corresponding village of the child has a private school; bank; cemented road and a private health clinic. Appendix table A1 gives detailed definition of each of these variables. Appendix table A2 presents the mean values of the outcome and control variables for the girls and boys separately and also tests whether the difference is statistically significant at 5% level.

---

[5] These tools can be accessed from http://www.asercentre.org/p/141.html (accessed on July 13, 2020)

[6] In India, a child of 6 years generally gets admitted to the first standard and hence by 8 years he/she is expected to complete the second standard.

Despite the considerable sample size and above mentioned these variables, one dimension where the ASER dataset lacks is information on several important indicators that might confound our results. For instance, social groups in terms of caste affiliation or religion might be important in our analysis to assess gender gap as heterogeneous effects within these groups may drive our population estimates. To overcome this, we use data from the IHDS, which has been conducted jointly by National Council of Applied Economic Research (NCAER) and University of Maryland in 2011-12. The dataset covered over 40,000 households gathered data on education, health, economic wellbeing, social status, and various other domains. Short tests capturing learning outcomes on reading, math and writing for children aged 8-11 years were also administered in the survey. These simple tests were administered to over 11,500 children (over 8000 children belonged to rural households) at their homes. Notably, these are the same tests are used by Pratham (a non-governmental organisation) which conducts assessment tests across the country for the ASER. For analysis with IHDS dataset as well, we use the same method to calculate the standardized score with a mean of one and zero standard deviation.

As in the earlier case, the main explanatory variable of interest is the gender of the child, which would indicate gender based differences arising independently after controlling for other factors. In terms of the covariates, this dataset allows us to use similar child and household level variables from IHDS that have been used in the analysis using the ASER data. These include age of the child, cemented house that is reflected through the type of household wall (whether concrete or not), possession of television, toilets and mobiles and usage of computers along with household size. In addition to these, we are able to include a number of relevant controls to the

Indian context that may affect learning outcomes - birth order; caste and religion; age, gender and level of education of household head; along with household level economic factors like yearly per capita consumption expenditure have been added as controls.

Further, school management (private or government run schools) is also controlled for following a number of studies which have seen the impact of attendance in private schools for the children on their learning outcomes (Chudgar and Quin 2012; Muralidharan and Sundararaman 2015; Singh 2015; Singhal and Das 2019). Other independent variables include whether the child has suffered from short-term illness or fever in the last 30 days prior to the survey, whether the distance from the household to the school is more than 1 kilometre and gender of the teacher. We have also added education of the mother and whether she is involved in household chores or other activities as control variables in the regressions. A significant factor that could deter spending time with school related activities is that girls could be spending more time cooking at home or doing chores within the house which has not been captured directly. Nevertheless, in the regression model, we also control for direct inputs like time spent in school, private coaching and for doing homework, which should serve as suitable proxies.[7]

To study the implications of societal gender norms and attitude on the score difference, we use data from the fourth wave of the National Family and Health Survey (NFHS-4). The survey was conducted in 2015-16 by the Ministry of Health and Family Welfare of The Government of India. 601,509 households across 640 districts were surveyed, which makes it representative not only at the national and the state level but also at the district level. Critical to the studied question, the survey provides information various indicators of female agency and attitude. We use district level estimates of the prevalence of these attitudes and norms and merge it with the ASER dataset collected during the same time period (2016) to examine if females from these districts fare systematically worse. To assess the implications of household norms, we use the first round of the IHDS dataset conducted in 2004-05 and surveyed largely the same households which were surveyed in the second round (IHDS-2). This allowed us to gauge the effects in households with rigid gendered norms that may have remained unaltered from 2004-05 to 2011-12.

---

[7] Appendix table A1 describes all these variables in details and appendix table A2 give the summary statistics for girls and boys separately.

*Estimation Strategy*

To examine the gender gap effects on learning outcomes, we firstly pool data from all the rounds of ASER starting from 2010 to 2018. The standardized reading and mathematics scores are taken as the main dependent variables. Whether the child is female or not becomes our variable of interest. Accordingly, we use the following regression:

$$Y_{ijst} = \alpha + \beta Female_{ijst} + \delta X_i + \gamma H_j + \sigma_t + \pi_s + \varepsilon_{ijs} \quad (1)$$

where $Y_{ijs}$ is the corresponding outcome variable in time, $t$ (2010 to 2018), for child, $i$ from household, $j$ situated in state, $s$. $Female_{ijs}$ is the dummy variable that takes the value of 1 if the child is female and 0 otherwise. $X_i$ is the vector of individual characteristics, $H_j$ is the vector of household and village characteristics and $\pi_s$ are state level dummy variables. $\sigma_t$ are the time fixed effects, which are dummy variables for each of the years of survey and the error term is given by $\varepsilon_{ijs}$. We use Ordinary Least Squares (OLS) for estimation with the standard errors clustered at the district level to account for the within-district heterogeneity. $\beta$ is our variable of interest. To gauge the effects from the IHDS 2011-12 dataset, we use the same model but with a much elaborate set of control variables captured by $X_i$ and $H_j$.

To estimate the potential effects across time to study whether there is an improvement in the gender gap from 2010 till 2018, we introduce an interaction term of the $Female_{ijst}$ and year dummies in the regressions using ASER data. The regression equation is as follows, whereby we analyse the $\theta$s:

$$Y_{ijst} = \alpha + \beta Female_{ijst} + \sum_{t=2010}^{2018} \theta_t (Female_{ijst} * \sigma_t) + \delta X_i + \gamma H_j + \sigma_t + \pi_s + \varepsilon_{ijs} \qquad (2)$$

To estimate the effects across age, an interaction term of the $Female_{ijst}$ and age dummies is introduced in equation (2) instead of the interaction term of the $Female_{ijst}$ and year dummies. This enables to us study whether there is a divergence or convergence in gender gap with age of the child.

## IV. RESULTS

*Analysis with ASER data*

As discussed, first using pooled data from 2010 to 2018, we examine the prevalence of gender gap in reading and mathematics learning, controlling for the potential confounders. Accordingly, we use equation (1) to estimate this and the marginal effects from the OLS regressions are presented in Table 1. Here we use three different specifications to ensure that our findings are robust to addition of other covariates. In the first specification, the child and the household characteristics are controlled for. In the second one, the village level characteristics are added and then in the third, which is our preferred specification, we incorporate the state level dummies as well along with the individual, household and village level factors. In all these specifications, survey round dummies are added to control for year specific difference that might be systematically correlated with gender of the child. The findings indicate no significant gap in terms of reading outcomes implying an average girl are equally likely to score well if compared to a similar boy child. However, with respect to mathematics, we observe a statistically significant difference with the female children much less likely to score as much as the boys. An average female child is found to score close to 0.08 standard deviations lower in comparison to a similar male child and this difference is found to be statistically significant at 1% level.

[Table 1 here]

*Analysis with IHDS*

The results from the regression outlined in equation (1) from the IHDS survey data is given in table 2. Here, we use three specifications. In the first one, we incorporate all the household and child level covariates that were included in the analysis with ASER. In the second specification, we add other variables that include other household level and school characteristics except the time spent on direct inputs like schooling, private coaching and for homework. In the third specification, which is the preferred one, we include these as well as crude substitutes for time use patterns. In all these specifications, state fixed effects are included and the standard errors are clustered at the district level. As in the earlier case, we find limited gender gap in reading abilities across all specifications. However, we observe female children are significantly more likely to score lesser in mathematics. In terms of the effect size, this is close to what we observe from the ASER data: females score about 0.1 standard deviation lower than that for the male children on average and this is robust across specifications including the one where we have controlled for household, child level and school characteristics.

[Table 2 here]

Are these differences observed across different household and child characteristics? The elaborate information available in the IHDS dataset allows us to decipher the gap across heterogeneous household groups and examine if the gender gap in mathematics learning remains pervasive.

*(a) Private and Government Schools[6]*

We extend our analysis further to explore if females perform better than the male children across private or government run schools. Figure 1a shows the marginal effects and the 95 per cent confidence intervals (CI) from the regressions that include all the mentioned covariates separately for sample of children studying in private and government run schools. We observe significantly lower mathematics scores among females in comparison to the males for both these regressions. Such, though is not the case for reading scores.

[Figure 1a here]

*(b) Social groups*

Social groups which categorize households into different caste and religion constitute an important dimension in the Indian society. Literature has indicated households belonging to backward castes including Scheduled Caste (SC), Scheduled Tribe (ST) and Muslim religion enjoy lowest autonomy, suffer from deprivation, discrimination and subsequently face inequality in opportunities in terms of health, education and employment (Thorat and Neuman, 2012; Banerjee, Duflo, Ghatak, and Lafortune 2013). Accordingly, we run separate regressions for children belonging to the following:

(i) *Brahmin* caste, Christian religion and other forward caste

(ii) Muslim and 'Other Backward Classes' (OBC)

(iii) *Dalits* (Scheduled Castes) and *adivasis* (Scheduled Tribes)

Individuals belonging to the first group are, on average, economically and socially better off as compared to those belonging to the other two groups. *Dalits* and *adivasis* form arguably the worst-off group both socially as well as economically and lag behind the upper castes in multiple indicators of welfare (Sundaram and Tendulkar 2003).

Figure 1b presents the marginal effects from the regressions run separately for children belonging to these three groups. Our findings give evidence of a persistent gender gap in mathematics learning outcomes within all these groups. In terms of reading though, no significant difference is found.

[Figure 1b here]

*(c) Economic groups*

Difference in mathematics scores across gender might be heterogeneous across households of different economic status. Apart from the prevalence of an apparent son preference as a choice, in poorer households parents may choose to invest more in boys because of higher economic returns (Rosenblum, 2017). Since numeracy or mathematical skills can be perceived to be important pre-requisite in the labour market, investment might be disproportionately higher for boys than girls especially when resources are scarce.

We test the presence of such heterogeneous association of girls scoring lesser in mathematics across these different economic groups. For this purpose, we divide the household yearly per-capita consumption expenditure into three equally divided terciles and run separate regressions with children from each of these groups of households. We consider consumption expenditure as it is arguably one of the best indicators of measuring economic well-being and has been used to estimate official poverty levels in India (Planning Commission, 2014). Figure 1c presents the marginal effects from the regressions. As in the other cases, we find girls from poorer households score significantly lesser on average than male children, which is found to be true even among children from the richest set of households.

[Figure 1c here]

*(d) Birth order*

Literature indicates various hypotheses about the impact of birth order on educational expenditure and achievements of children. Those predicting a negative hypothesis suggest reasons such as greater parental involvement and responsibility towards children of lower birth order. The parents also get older when they rear children of higher birth order. However, those predicting a positive relationship put forward reasons like growth of family income over the life cycle, experience gained by parents towards child rearing and assistance provided by the older children in terms of finance and caring (Booth and Kee 2009). Accordingly, we examine if the gender gap in mathematics scores is prevalent among those of different birth orders. For this purpose we categorize children into two groups: those with birth order '1' and those with birth order '2' or above. Figure 1d presents the marginal effects from the OLS regressions across these two groups of children.

Our findings appear to suggest that mathematics scores for girls across birth order 1 or 2 or above is lower than that for boys on average and the difference is statistically significant. Notably as discussed, sex selection is found less likely to be prevalent among children of first birth order, which is why the gender of these children might be assumed as close to exogenous. The fact that the gender gap is significant among children of the first birth order confirms that our findings are robust and not confounded by a potential selection bias against female children in the pre natal stage.

[Figure 1d here]

*Changes over time*

Next, we examine the changes in the gender gap since 2010 and whether we get any indication of an improvement through reduction in gender gap. For this, we run regression using the equation outlined in (2). The marginal effects for reading and mathematics outcomes are presented in figure 2(a) and 2(b) respectively. Interestingly we find a reverse gender gap in reading scores that started in 2014 and the difference in 2018 is found to be statistically significant at 5% level. This implies girl students have been faring better, on average, than their male counterparts since the last five years on reading skills. Yet, when we consider mathematics scores, we find a gender gap biased against the female children and the effects seem to be persistent across the years starting from 2010 to 2018.

[Figure 2a and 2b here]

*Changes over age*

Next, we assess the convergence or divergence in the gender gap among the children with age. The plots for the marginal effects on reading outcomes which are presented in figure 3(a) indicate prevalence of no gender gap across the whole range of age starting from 8 years to 16 years. In other words, controlling for other factors, a girl, irrespective of her age, is likely to perform similarly to that of boy of same age. Nevertheless, as found in the last section, we observe a discernible gender gap for mathematics, which starts diverging with age (figure 3(b)).

[Figure 3a and 3b here]

Please note that in the above regression, we have controlled for enrollment status of the child in school. This ensures we are controlling for the prevalent gender gap that starts intensifying with age and hence our findings are unlikely to be confounded by lesser female enrollment at higher age. In other words, because literature has documented higher school drop-out among females with age, it is more likely that a boy child with similar innate ability to a girl child may remain in school and gain better learning capabilities in comparison to the latter with respect to secondary schooling. However, the results do not change even if we only consider the enrolled children.

We test this in another way where we consider the children who were of age 8 years in 2010 and tracked them till 2018. Accordingly, we include only children of age 9 years in 2011, 10 years in 2012, 11 years in 2013, 12 years in 2014, 14 years in 2016 and 18 years in 2018. This gives an opportunity to examine the changes in gender gap as we move through similar set of children from 2010 to 2018. Therefore, we are able to gauge if the gender gap at an older age (14 or 16 years) increases even if there is no gap among similar set of children when they were 10 years old. For this, we create a *time* variable which takes the value of 1 for children surveyed in 2010, 2 for those surveyed in 2011, 3 for 2012 and so on till 9 in 2018 and run a regression for the sample of children as given above. Here we use equation (2) but with an interaction term of the $Female_{ijst}$ and time dummies and examine the relevant coefficients. We find no change in reading score across age but a significant divergence in mathematics score at higher age is observed. This is despite the fact that the gender gap at lower age is statistically insignificant at

5% level. Supplementary figure SF1 and SF2, which present the marginal effects on reading and mathematics scores also indicate the same.

*State level analysis*

One of the key features of India is the huge variation in levels of development across states. Research indicates considerable regional inequality in the pace of economic growth and poverty reduction (Nagaraj, 2000; Ahluwalia, 2002). Findings largely indicate North Indian states like Bihar and Uttar Pradesh show low levels of economic growth whereas western states like Gujarat, Maharashtra and the south Indian state of Tamil Nadu showed increased growth in the 1990s. In addition, with respect to poverty reduction and human development indicators, evidence seems to suggest prevalence of substantial inequality across the Indian states (Deaton and Dreze, 2002; Ravallion and Datt, 2002; Purohit, 2004; Purohit, 2008). This appears to be evident with regards to gender inequality as a recent study by the Government of India indicates that the South Indian states of Kerala, Andhra Pradesh and Karnataka are found to be performing better while states like Uttar Pradesh, Bihar and Madhya Pradesh among others lag behind. Putting everything together, existing studies indicate southern Indian states to be developed with higher levels of empowerment among women (Jejeebhoy and Sathar, 2001; Evans, 2020).

In this context, we use the ASER data to generate state level estimates to examine the states which have higher levels of gender gap and those with lower levels or even show a reverse gap in terms of mathematics abilities. Table 3 presents the main results from regressions run separately for the major states of India as elucidated in equation (1). Here, based on data from 2011 to 2018, we categorize the states into three groups: states with significant female gender gap (girl disadvantage), (ii) states with reverse gender gap (boy disadvantage) and (iii) states with insignificant gap (no disadvantage). Expectedly, we see a regional pattern emerge in the data. We find that 12 out of the 29 states show female disadvantage which is statistically significant at 1% on average.

This comprises northern and central Indian states that include Rajasthan, Madhya Pradesh, Uttar Pradesh, Bihar, Chhattisgarh and Jharkhand along with eastern states like Orissa, West Bengal and Assam. Southern Indian states that include Karnataka, undivided Andhra Pradesh, Tamil Nadu, Kerala and Pondicherry show male disadvantage as compared to the females (reverse gender gap). Major western states that include Maharashtra and Gujarat show no statistically significant girl disadvantage. This is true for most of the North-Eastern states like Arunachal Pradesh, Mizoram, Nagaland and Sikkim.

[Table 3 here]

Next, we use equation (2) to examine the changes in the gender gap temporally from 2010 to 2018 for these states. Figure 4 presents these estimations over time for the five states (Bihar, Uttar Pradesh, Madhya Pradesh, Rajasthan and Jharkhand) which show a more pronounced trend of increasing gender gap. In the other states that include Assam, Orissa, West Bengal and Chhattisgarh which have a female disadvantage on average, there are signs of improvement (Appendix figure F1). Notably, Tamil Nadu appears to be a state where we find an increase trend of reverse gender gap. This indicates systematically higher boy disadvantage in the recent years which needs immediate recognition and implementation of relevant policies. Nevertheless, in states that show reverse gender gap on average, we find no evidence of temporal disadvantage for boys. Surprisingly, Punjab stands out in this analysis since there is a significant gender gap in education spending in Punjab along with states that include Bihar, Rajasthan, Madhya Pradesh and Uttar Pradesh (Kingdon 2005). Further research is needed to understand this issue in greater detail

[Figure 4 here]

*Implications of gender norms*

Our state wise analysis indicates two important facets: first, there is a wide variation in the extent of gender gap in mathematics across states and second, there exist a systematically perverse female disadvantage especially in northern states which is in fact increasing over time. Research indicates these states are massively patrilineal with the female biased cultural norms and attitudes being deeply rooted among large sections of the population (Dyson and Moore, 1983). In fact it is often observed that these regressive gender norms exist not only in households but also within the society more broadly. In fact, literature in varied context have established how gender in mathematics correlates with male preference within the family as well as societal gender norms (Guiso et al. 2008; Pope and Sydnor, 2010; Nollenberger et al. 2016; Dossi et al. 2021). In this section we formally examine if pre-existing gender norms at household level is in anyway correlated with the prevalent gender gap.

For this, we use the IHDS data collected in 2011-12 and utilize two possible indicators of gender stereotyping emanating from patriarchal cultural norms and traditions. Seclusion practices in households such as *purdah* or *ghunghat* (the practice of women veiling their faces) are examples of such norms or prejudice. Among Hindu households, the practice of *ghunghat* is highly prevalent in Northern India (less prevalent in Southern parts), while purdah is prevalent in Muslims across India, and those practicing this are found to have lesser autonomy in going outside the house to work or meet friends, take household decisions or participating in the labour market (Rahman and Rao, 2004; Coffey, Hathi, Khurana and Thorat 2018).

Another form of such prejudice faced by women is the practice of them eating food after men which have been associated with the likelihood of women being underweight putting them at higher health risks (Coffey et al., 2018). During pregnancy, this can also translate into adverse outcomes for their children (Coffey, 2015). Since these norms and attitudes are patriarchal in nature, it is likely that in households where such these gendered norms or prejudices are followed, gender stereotyping would be higher.

To gauge whether female children from households where these norms are practiced (*ghunghat* and the practice of women eating food after men) are likely to score worse in mathematics, we run similar regression as elucidated in equation (2) with an interaction term of the $Female_{ijst}$ and these indicators instead of the female-time interaction. For this we also make use of the longitudinal IHDS-1 data that collected similar information from the same set of households in 2004-05 (Desai et al., 2010).[8] Figure 5 presents the regression results to assess the gender gap in households that follow these practices. In the first set of regression, we examine the gap in households that practiced these in 2004-05, in the second set, we assess the gap in households practicing these in 2011-12 and then in the third, we study this in households who reported of having practiced these both in 2004-05 and 2011-12. Across these regressions, we find that the mathematics scores among females to be systematically lower in households practicing these regressive gender norms.

[Figure 5 here]

We further explore whether societal norms are associated with the gender gap. Here, we make use of district representative data from the National Family Health Survey (NFHS) conducted in 2015-16. In particular, we estimate gender norms and attitude at district level and

---

[8] Please refer to https://ihds.umd.edu/ for more details (accessed September 28, 2017)

merge that with the ASER data for 2016 to assess if female children from districts with higher prevalence of regressive gendered norms and attitudes score worse in mathematics relative to the males. We consider a number of district level indicators (in terms of proportion) that capture female agency not only with respect to their treatment within the households but also their bargaining power:

(i) Husband/ partner jealous if respondent talks with other men

(ii) Husband/ partner accuses respondent of unfaithfulness

(iii) Husband/ partner does not permit respondent to meet female friends

(iv) Husband/ partner tries to limit respondent's contact with family

(v) Husband/ partner insists on knowing where respondent is

(vi) Husband/ partner doesn't trust respondent with money

(vii) Females ever been humiliated by husband/partner

(viii) Females ever been threatened with harm by husband/partner

(ix) Females been insulted or made to feel bad by husband/partner

(x) Females who usually decide on respondent's health care

(xi) Females who usually decide on large household purchases

(xii) Females who usually decide on visits to family or relatives

(xiii) Females who usually decide what to do with money husband earns

Note that for indicator (i)-(ix), higher values imply gender regressive norms while a lower value for (x)-(xiii) implies that.

With this, we run regression to assess if females from district with higher prevalence of gender insensitive practices are less likely to score well in mathematics. Figure 6 show the marginal effects from interaction of the $Female_{ijst}$ dummy and these indicators as mentioned earlier. The findings indicate a strong association with the female children performing systematically worse on average than the male children from districts with regressive norms. Importantly, we find these to be true not only for indicators which capture respect or the females within the household (indicator i-ix) but also those that indicate their decision making power (indicator x-xiii). For these indicators, we find that in districts with higher prevalence of females taking household decisions, an average girl child is likely to score better in mathematics compared to a similar boy child.

[Figure 6 here]

## V. DISCUSSION AND CONCLUSION

This paper provides an overview on the prevalence and persistence of gender gaps in mathematics among adolescent children from rural India across age, time and geographical location. We utilize multiple datasets that include the over 2 million children from the ASER survey data from 2010 to 2018 apart from the longitudinal IHDS survey conducted in 2004-05 and 2011-12. We find a robust and strong evidence of a gender gap in basic mathematical abilities that appear to be prevalence across adolescent females across economic strata, social groups and various other demographic compositions. Worryingly, the gap persists across age groups and has been increasing temporally on average. We also find significant inter-state variation with most of the Northern states, where we observe a disproportionately larger gap.

One of the important dimensions of the inter-state variation is the southern states in particular exhibit a reverse gender gap (where boys are performing worse than girls). While this reverse gap is unexpected and requires further research, the lack of female disadvantage for these outcomes in southern states is unsurprising because such regional differences are common due to variation in historical contexts, norms, structures and relevant investments made in these states. Evans (2020) highlights higher segregation and lower labour force participation of females in northern India. In addition, there are matrilineal structures in southern India due to which certain norms associated with these differences in particular may be limited. In comparison, Singh et al. (2021) find higher levels of patriarchy in northern and western states relative to the southern states. Accordingly, we complement our understanding regarding the association of gender based patriarchal norms with mathematics learning of females using an additional data source (NFHS), which constitute the dominant explanations for these differences in the nascent but growing literature on this topic. Expectedly, we observe a strong association of the pre-existing household and societal level patriarchal norms and attitudes with gender differences in mathematics ability.

Despite the evident scale and stakes of this problem, there is currently a lack of adequate recognition and effective policies that have targeted (or been successful) in improving these gaps. This is despite the formal recognition of such differences and conjecture on potential causes that goes as back as 2005 as part of the National Curriculum Framework (NCERT, 2005) which is commissioned by National Council for Educational Research and Training (NCERT), one of India's apex education bodies. While there exist a number of policies and programmes targeting to improve low learning levels of all children that seek to address female disadvantage in education, these are not meant to exactly redress the specific issue of differences in numeracy levels. NCERT (2005) along with other studies have pointed out that gendered learning material and behaviour of the teacher and parents (for example, attributing better performance in mathematics of boys to 'intelligence' and of girls to 'hard work') may play a role in reducing the self-efficacy of the girl child. The fact that the gender gap has increased from 2010 to 2018 indicates that the current gamut of policies has not been able to effectively target. However as southern States show a reverse gender gap, it may require a different set of policies altogether and might be caused by different reasons, something that future studies can explore

There is growing discourse on ensuring foundational literacy and numeracy (FLN) among children, as part of the recent National Education Policy in India. Even though this too does not acknowledge these gaps explicitly, it is an important step towards the formal recognition of learning deficits of children highlighted each year by the annual ASER reports. With states in the process of ratifying the new policy and bringing in legislations to implement this, there is an opportunity to set up processes to build capacity to conduct periodic assessments of all children (across grades). While some states have started building these capabilities already, it will be essential to ensure that these assessments to generate actionable insights that can inform policy interventions and responses at state or a more disaggregated level.

Table 1: Estimation of reading and mathematics scores from ASER dataset

|  | Reading scores | | | Mathematics score | | |
|---|---|---|---|---|---|---|
|  | (1) | (2) | (3) | (4) | (5) | (6) |
| Female child | 0.004 | 0.003 | 0.002 | -0.078*** | -0.078*** | -0.079*** |
|  | (0.004) | (0.004) | (0.004) | (0.005) | (0.005) | (0.005) |
| Year FE | Yes | Yes | Yes | Yes | Yes | Yes |
| Individual and household characteristics | Yes | Yes | Yes | Yes | Yes | Yes |
| Village characteristics | No | Yes | Yes | No | Yes | Yes |
| State FE | No | No | Yes | No | No | Yes |
| R-square | 0.264 | 0.265 | 0.275 | 0.274 | 0.275 | 0.294 |
| Observations | 2,241,960 | 2,241,960 | 2,241,960 | 2,235,782 | 2,235,782 | 2,235,782 |

*Note: Marginal effects from OLS regression are presented with the standard errors, clustered at the district level given in the parenthesis. The dependent variable is standardized reading and mathematics scores. The sample includes children in the age group 8-16 years from ASER 2010 to 2018. \*\*\* p<0.01, \*\* p<0.05, \* p<0.1. Estimation results for all the other variables are given in Supplementary table ST1.*

Table 2: Estimation of reading and mathematics scores from IHDS dataset

|  | Reading scores | | | Mathematics score | | |
|---|---|---|---|---|---|---|
|  | (1) | (2) | (3) | (4) | (5) | (6) |
| Female child | -0.036 | -0.030 | -0.040* | -0.113*** | -0.102*** | -0.108*** |
|  | (0.022) | (0.021) | (0.022) | (0.021) | (0.021) | (0.022) |
| Individual and household characteristics | Yes | Yes | Yes | Yes | Yes | Yes |
| Household and other school characteristics | No | Yes | Yes | No | Yes | Yes |
| Time use characteristics | No | No | Yes | No | No | Yes |
| State FE | Yes | Yes | Yes | Yes | Yes | Yes |
| R-square | 0.168 | 0.295 | 0.300 | 0.199 | 0.303 | 0.313 |
| Observations | 8,124 | 7,286 | 6,663 | 8,089 | 7,258 | 6,642 |

*Note: Marginal effects from OLS regression are presented with the standard errors, clustered at the district level given in the parenthesis. The dependent variable is standardized reading and mathematics scores. The sample includes children in the age group 8-11 years from IHDS 2011-12.  \*\*\* p<0.01, \*\* p<0.05, \* p<0.1*

Table 3: Gender gap across states

| States | Marginal effects | 95% Confidence Interval | | P-value | Observations |
|---|---|---|---|---|---|
| *Girl Disadvantage* | | | | | |
| Assam | -0.049 | -0.065 | -0.032 | 0.000 | 79439 |
| Bihar | -0.170 | -0.181 | -0.160 | 0.000 | 190637 |
| Chhattisgarh | -0.056 | -0.077 | -0.036 | 0.000 | 58963 |
| Jammu & Kashmir | -0.103 | -0.137 | -0.069 | 0.000 | 40762 |
| Jharkhand | -0.106 | -0.126 | -0.087 | 0.000 | 90762 |
| Madhya Pradesh | -0.108 | -0.130 | -0.085 | 0.000 | 187580 |
| Manipur | -0.048 | -0.074 | -0.021 | 0.002 | 31398 |
| Orissa | -0.050 | -0.071 | -0.030 | 0.000 | 98656 |
| Rajasthan | -0.140 | -0.162 | -0.118 | 0.000 | 138745 |
| Uttar Pradesh | -0.203 | -0.214 | -0.192 | 0.000 | 350467 |
| Uttaranchal | -0.115 | -0.135 | -0.094 | 0.000 | 44773 |
| West Bengal | -0.064 | -0.098 | -0.030 | 0.000 | 47367 |
| *Boy Disadvantage* | | | | | |
| Andhra Pradesh | 0.039 | 0.021 | 0.058 | 0.000 | *66319* |
| Karnataka | 0.045 | 0.030 | 0.061 | 0.000 | *106464* |
| Kerala | 0.113 | 0.096 | 0.129 | 0.000 | *38467* |
| Meghalaya | 0.065 | 0.044 | 0.086 | 0.000 | *22168* |
| Pondicherry | 0.101 | 0.055 | 0.146 | 0.006 | *5658* |
| Punjab | 0.107 | 0.084 | 0.130 | 0.000 | *58514* |
| Tamil Nadu | 0.086 | 0.074 | 0.099 | 0.000 | *100527* |
| Goa | 0.037 | 0.020 | 0.055 | 0.012 | *3526* |
| *No Disadvantage (at 5% level)* | | | | | |
| Arunachal Pradesh | -0.017 | -0.046 | 0.011 | 0.218 | *27459* |
| Gujarat | -0.010 | -0.027 | 0.006 | 0.228 | *92101* |
| Haryana | -0.029 | -0.061 | 0.004 | 0.080 | *78480* |
| Himachal Pradesh | 0.018 | -0.009 | 0.045 | 0.188 | *40476* |
| Maharashtra | 0.000 | -0.015 | 0.014 | 0.960 | *121100* |
| Mizoram | -0.002 | -0.028 | 0.025 | 0.898 | *32450* |
| Nagaland | -0.017 | -0.036 | 0.002 | 0.081 | *45427* |
| Sikkim | 0.007 | -0.010 | 0.024 | 0.370 | *10281* |
| Tripura | 0.001 | -0.038 | 0.040 | 0.943 | *12291* |

*Note: Confidence intervals are calculated by clustering the standard errors at the district level. The dependent variable is standardized mathematics scores. The sample includes children in the age group 8-16 years from ASER 2010 to 2018.*

Figure 1: Gender gap across heterogeneous households
(a) School types

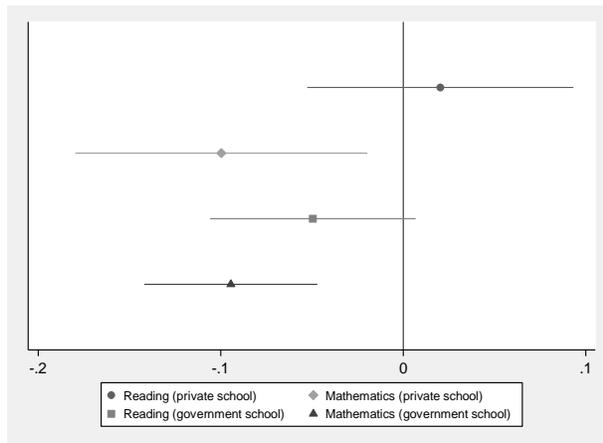

(b) Social groups

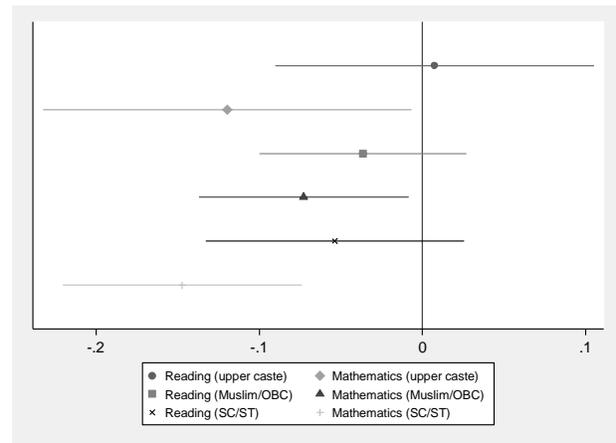

(c) Economic groups

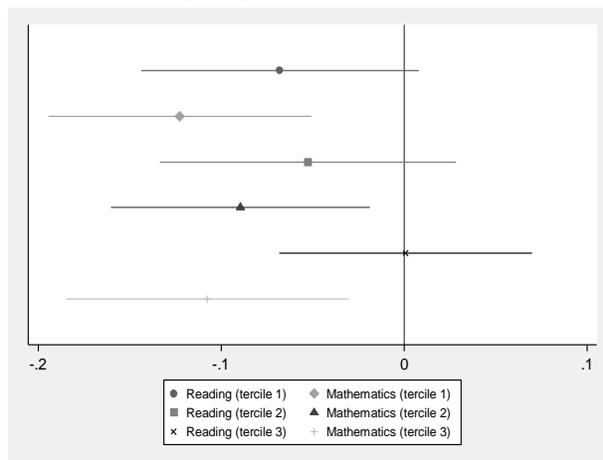

(d) Birth-order

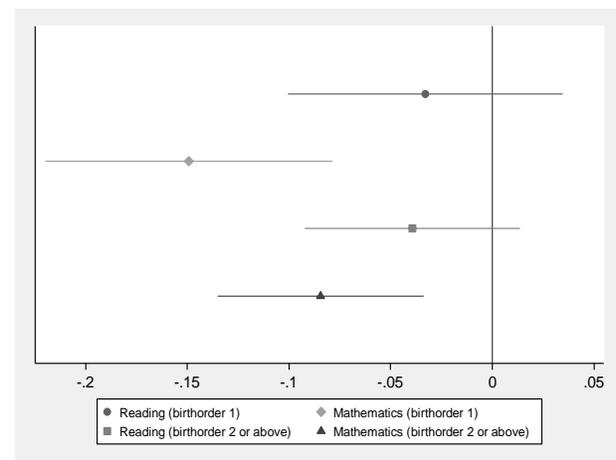

*Note: 95% Confidence intervals are calculated by clustering the standard errors at the district level are plotted along with the marginal effects. The sample includes children in the age group 8-11 years from IHDS 2011-12.*

Figure 2: Gender gap across time
(a) Reading scores

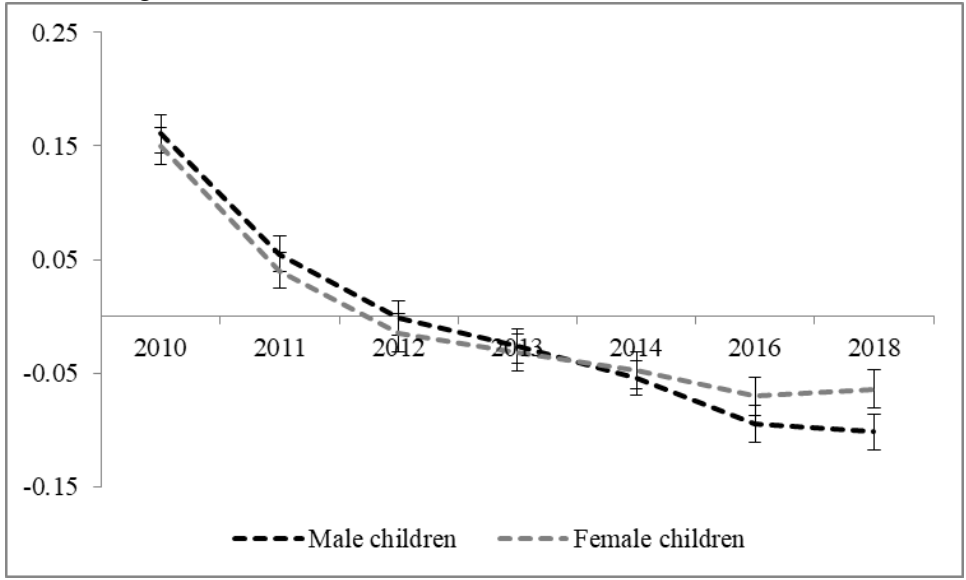

(b) Mathematics scores

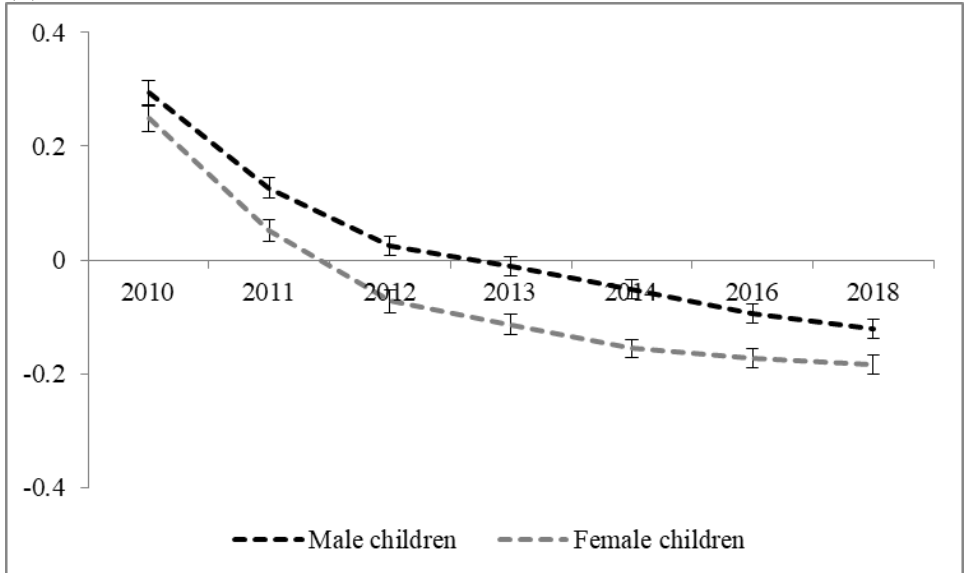

*Note: 95% Confidence intervals are calculated by clustering the standard errors at the district level are plotted along with the marginal effects. The sample includes children in the age group 8-16 years from ASER 2010 to 2018.*

Figure 3: Gender gap across child age
(a) Reading score

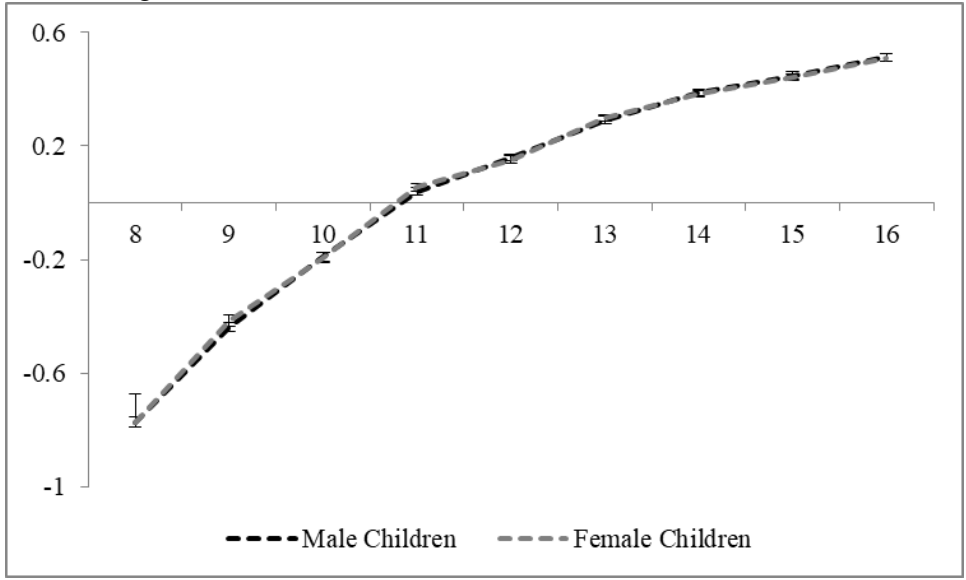

(b) Mathematics score

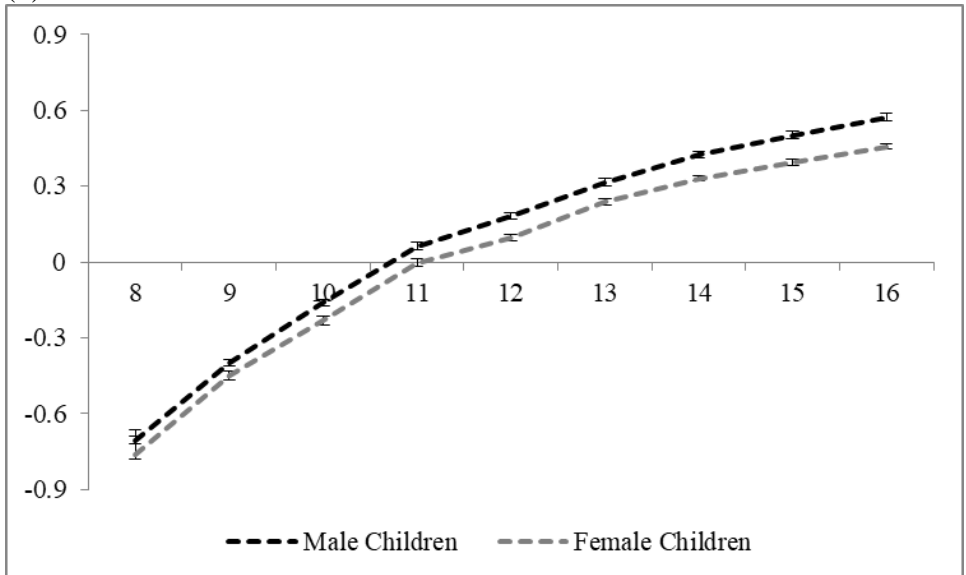

*Note: 95% Confidence intervals are calculated by clustering the standard errors at the district level are plotted along with the marginal effects. The sample includes children in the age group 8-16 years from ASER 2010 to 2018.*

Figure 4: Gender gap over time across states
(a) Bihar

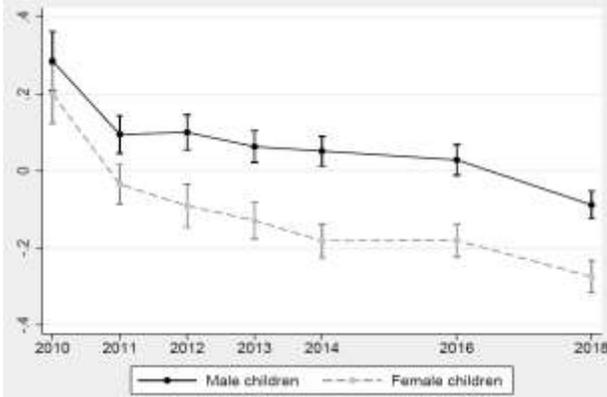

(b) Jharkhand

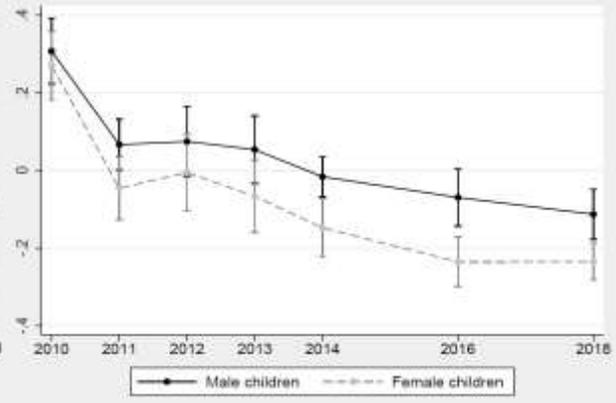

(c) Madhya Pradesh

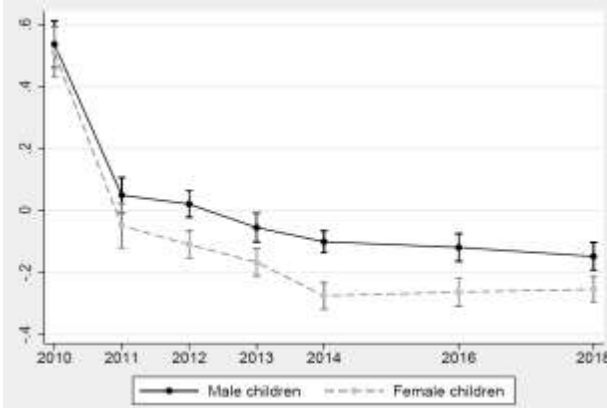

(d) Rajasthan

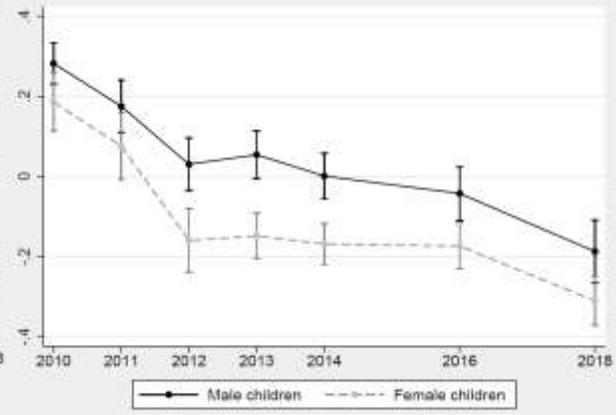

(e) Uttar Pradesh

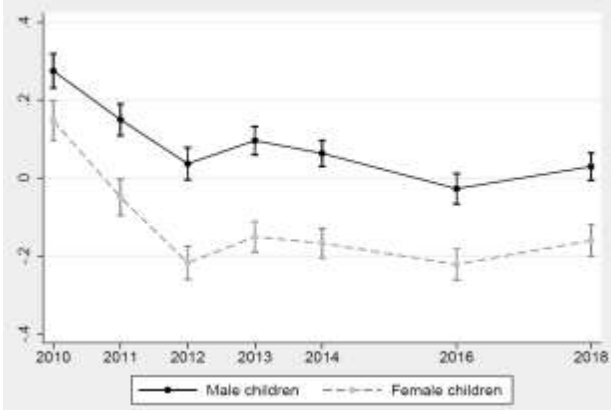

*Note: 95% Confidence intervals are calculated by clustering the standard errors at the district level are plotted along with the marginal effects. The sample includes children in the age group 8-16 years from ASER 2010 to 2018 from the respective states.*

Figure 5: Interaction effects

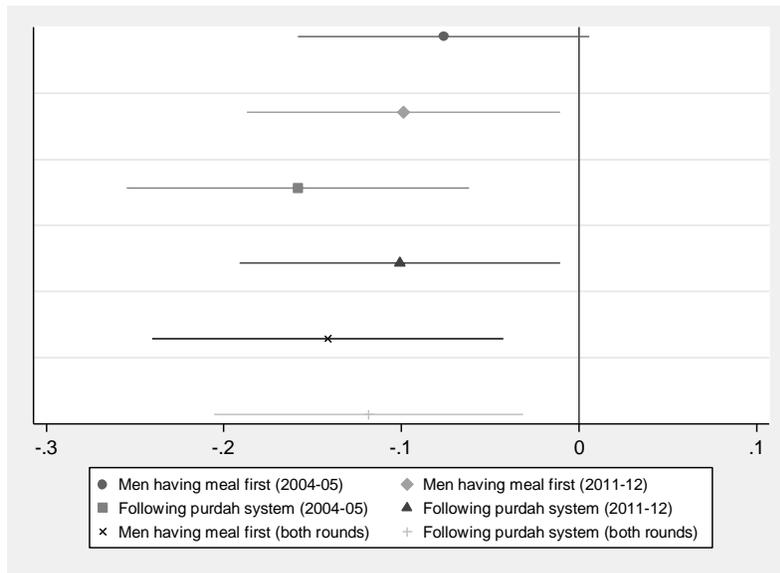

*Note: 95% Confidence intervals are calculated by clustering the standard errors at the district level are plotted along with the marginal effects. The sample includes children in the age group 8-11 years from IHDS 2011-12.*

Figure 6: Interaction with indicators from NFHS-4

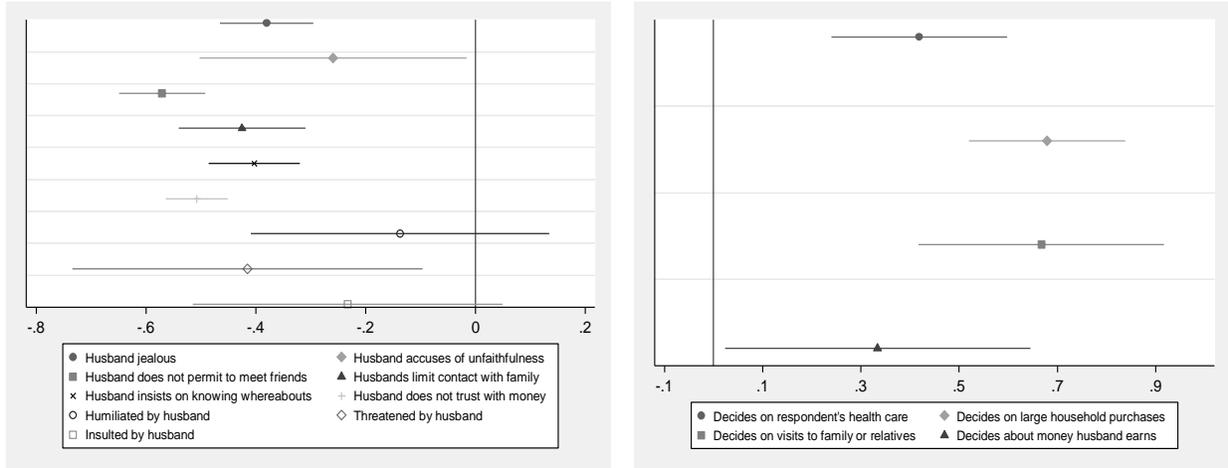

*Note: 95% Confidence intervals are calculated by clustering the standard errors at the district level are plotted along with the marginal effects. The sample includes children in the age group 8-16 years from ASER 2016.*

Appendix table A1: Variables used and the source

| Variables | Definition |
|---|---|
| *Source: ASER (2010-18)* | |
| Standardized reading scores | Standardized value of reading scores with the following five categories: cannot read anything (1); can identify letters (2), can read words (3), can read a standard 1 level text (4), and can read a standard 2 level text (5) |
| Standardized math scores | Standardized value of mathematics scores with the following five categories: cannot do any arithmatic (1); recognition of number 1-9 (2), recognition of numbers 11-99 (3), can do a two-digit subtraction with a carry over (4), and can do a division (three by one form) division (5) |
| Gender of the child | 1 if the child is female; 0 otherwise |
| Age of the child | Age of the child in years |
| Enrolled in school | 1 if the child is enrolled in school; 0 otherwise |
| Resides in cemented house | 1 if the walls of the house is made of bricks and cement (*pucca*); 0 if *semi-kutcha* or *kutcha* |
| Has a toilet | 1 if there is a toilet in the household; 0 otherwise |
| Has a television | 1 if there is a television in the household; 0 otherwise |
| Knows computer | 1 if anybody in the household knows computer; 0 otherwise |
| Has a mobile phone | 1 if there is a mobile phone in the household; 0 otherwise |
| Household size | Total members in the household who eat from the same kitchen |
| Village has cemented road | 1 if the sampled village has tarred metal road leading to it; 0 otherwise |
| Village has post office | 1 if the sampled village has a post-office; 0 otherwise |
| Village has bank | 1 if the sampled village a bank; 0 otherwise |
| Village has government primary health clinic | 1 if the sampled village has government primary health clinic; 0 otherwise |
| Village has private health clinic | 1 if the sampled village has a private health clinic; 0 otherwise |
| Village has private school | 1 if the sampled village has a private school; 0 otherwise |
| *Source: IHDS (2011-12)* | |
| Standardized reading scores | Standardized value of reading scores with the following five categories: cannot read at all (=0), can recognise letters but not words (=1), can read words but not a paragraph (=2), can read a paragraph but not a story (=3) and can read a story (=4) |
| Standardized math scores | Standardized value of mathematics scores with the following four categories: unable to recognise numbers (=0), recognise numbers but are unable to do arithmetic (=1), can do a subtraction problem but not division (=2) and can solve a division problem (=3) |
| Age of the child | Age of the child in years |
| Resides in house with concrete walls | 1 if the predominant wall of the household is concrete/ cemented; 0 otherwise |
| Has a toilet | 1 if there is a toilet in the household; 0 otherwise |
| Has a television | 1 if there is a television in the household; 0 otherwise |
| Knows computer | 1 if anybody in the household knows computer; 0 otherwise |
| Has a mobile phone | 1 if anybody in the household uses mobile phone; 0 otherwise |
| Household size | Total members in the household who live under the same roof and share the same kitchen for 6+ months in the year prior to the survey |
| Brahmins, Other Forward Castes, Jains, Christians and Sikhs | 1 if the household belongs from Brahmin/ other forward caste/ Jains/ Christians or Sikhs; 0 otherwise |
| Muslims | 1 if the household belongs from Islam religion; 0 otherwise |
| Other Backward Classes | 1 if the household belongs from Other Backward Caste (OBC); 0 otherwise |
| Scheduled Castes | 1 if the household belongs from Scheduled Castes (SC); 0 otherwise |
| Schedule Tribe (Adivasi) | 1 if the household belongs from Scheduled Tribes (ST) or adivasi; 0 otherwise |
| Birth order | Birth order of the child |
| Grade | School grade in which th child is study |
| Short-term morbidity | 1 if the child had either fever, cough or diarrhea in the last 30 days; 0 otherwise |
| Attends government school | 1 if the child is enrolled in government school; 0 otherwise |
| Distance to school | Distance of the school from household (in km) |
| Teacher gender: Female | 1 if the gender of the class teacher is female; 0 otherwise |
| Yearly per capita expenditure of household (Indian rupees) | Yearly per capita consumption expenditure of the household (in Indian rupees) |
| Homework hours/week | Number of hours spent per week by child in the last 30 days for doing homework |
| School hours/ week | Number of hours spent per week by child in the last 30 days in school |
| Log of tuition hours/week | Logarithmic value of number of hours spent per week by child in the last 30 days in private tuition plus 1. |
| Age of household head | Age of the household head in years |
| Female head | 1 if the gender of the household head is female; 0 otherwise |
| No Education (head's education) | 1 if household head has received no education; 0 otherwise |
| Up to 8th grade (head's education) | 1 if household head has received upto 8 grade education; 0 otherwise |
| Above $8^{th}$ grade (head's education) | 1 if household head has received more than $8^{th}$ grade education; 0 otherwise |
| Mother's age | Age of the mother in years |
| No Education (mother's education) | 1 if mother has received no education; 0 otherwise |
| Up to 8th grade (mother's education) | 1 if mother has received upto 8 grade education; 0 otherwise |
| Above $8^{th}$ grade (mother's education) | 1 if mother has received more than $8^{th}$ grade education; 0 otherwise |
| Engaged in work outside home | 1 if mother in engaged in working outside; 0 otherwise |
| Following purdah system | 1 if anybody in the household practices ghunghat/ burkha/ pallu/ purdah; 0 otherwise |

| Men having meal first | 1 if men in the family eat first while taking the main meal; 0 otherwise |
|---|---|
| *Source: IHDS (2004-05)* | |
| Following purdah system | 1 if anybody in the household practices ghunghat/ burkha/ pallu/ purdah; 0 otherwise |
| Men having meal first | 1 if men in the family eat first while taking the main meal; 0 otherwise |
| *Source: NFHS (2015-16)* | |
| Husband jealous | District level proportion of women whose husband/ partner are jealous if they talk to other men |
| Husband accuses of unfaithfulness | District level proportion of women whose husband/ partner accuses them of unfaithfulness |
| Husband does not permit to meet friends | District level proportion of women whose husband/ partner does not permit them to meet female friends |
| Husband limits contact with family | District level proportion of women whose husband/ partner tries to limit their contact with family |
| Husband insists on knowing whereabouts | District level proportion of women whose husband/ partner insists on knowing where they are |
| Husband doesn't trust with money | District level proportion of women whose husband/ partner doesn't trust them with money |
| Humiliated by husband | District level proportion of women who have been humilitaed by their husbands/ partners |
| Threatened with harm by husband | District level proportion of women who have been threatened with harm by their husbands/ partners |
| Insulted by husband | District level proportion of women who have been insulted or made to feel bad by their husbands/ partners |
| Decides on respondent's health care | District level proportion of women who solely take decisions on her health care |
| Decide on large household purchases | District level proportion of women who solely take decisions on large household purchases |
| Decide on visits to family or relatives | District level proportion of women who solely take decisions on her visit to family or friends |
| Decide about money husband earns | District level proportion of women who solely can take decisions on spending what the husband/ partner earns |

Appendix table A2: Descriptive statistics (mean/proportion)

| | Total (1) | Boys (2) | Girls (3) | Difference (2)-(3) |
|---|---|---|---|---|
| *From ASER (2010-2018)* | | | | |
| Standardized reading scores | 0 | -0.001 | 0.001 | -0.00 |
| Standardized math scores | 0 | 0.040 | -0.043 | 0.083*** |
| Age of the child (in years)# | 11.746 | 11.725 | 11.77 | -0.045*** |
| Enrolled in school | 0.933 | 0.937 | 0.93 | 0.007*** |
| Resides in cemented house | 0.379 | 0.377 | 0.38 | -0.003*** |
| Has a toilet | 0.471 | 0.468 | 0.475 | -0.006*** |
| Has a television | 0.513 | 0.51 | 0.516 | -0.005*** |
| Knows computer | 0.154 | 0.156 | 0.152 | 0.004*** |
| Has a mobile phone | 0.745 | 0.744 | 0.746 | -0.002*** |
| Household size | 6.501 | 6.39 | 6.622 | -0.232*** |
| Village has cemented road | 0.786 | 0.785 | 0.787 | -0.002*** |
| Village has post office | 0.416 | 0.414 | 0.418 | -0.004*** |
| Village has bank | 0.254 | 0.252 | 0.256 | -0.005*** |
| Village has government primary health clinic | 0.421 | 0.42 | 0.421 | -0.002*** |
| Village has private health clinic | 0.315 | 0.318 | 0.313 | 0.005*** |
| Village has private school | 0.422 | 0.422 | 0.421 | 0.001 |
| *From IHDS (2011-12)* | | | | |
| Standardized reading scores# | 0 | 0.039 | -0.042 | 0.081*** |
| Standardized math scores# | 0 | 0.077 | -0.084 | 0.160*** |
| Age of the child# | 9.492 | 9.497 | 9.485 | 0.012 |
| Resides in house with concrete walls | 0.600 | 0.605 | 0.594 | 0.011 |
| Has a toilet | 0.357 | 0.367 | 0.347 | 0.020** |
| Has a television | 0.495 | 0.509 | 0.48 | 0.029*** |
| Knows computer | 0.019 | 0.023 | 0.014 | 0.010*** |
| Has a mobile phone | 0.196 | 0.217 | 0.173 | 0.044*** |
| Household size# | 6.551 | 6.391 | 6.725 | -0.334*** |
| Brahmins, Other Forward Castes, Jains, Christians and Sikhs | 0.172 | 0.183 | 0.16 | 0.023*** |
| Muslims | 0.133 | 0.128 | 0.138 | -0.01 |
| Other Backward Classes | 0.348 | 0.347 | 0.348 | -0.001 |
| Scheduled Castes | 0.234 | 0.232 | 0.236 | -0.004 |
| Schedule Tribe (Adivasi) | 0.113 | 0.109 | 0.117 | -0.008 |
| Birth order# | 2.217 | 2.208 | 2.227 | -0.018 |
| Grade# | 3.461 | 3.457 | 3.465 | -0.008 |
| Short-term morbidity | 0.185 | 0.184 | 0.187 | -0.003 |
| Attends government school | 0.659 | 0.614 | 0.707 | -0.093*** |
| Distance to school# | 0.235 | 0.263 | 0.205 | 0.058*** |
| Teacher gender: Female | 0.389 | 0.363 | 0.418 | -0.055*** |
| Yearly per capita expenditure of household (Indian rupees)# | 16958.88 | 17591.16 | 16275.33 | 1315.84*** |
| Homework hours/week# | 7.191 | 7.336 | 7.033 | 0.303*** |
| School hours/ week# | 32.572 | 32.449 | 32.707 | -0.258 |
| Log of tuition hours/week[#] | 1.58353 | 1.775 | 1.375 | 0.400*** |
| *Household head characteristics* | | | | |
| Age[#] | 45.975 | 46.051 | 45.893 | 0.158 |
| Female | 0.114 | 0.113 | 0.114 | -0.001 |
| No Education | 0.399 | 0.399 | 0.4 | -0.001 |
| Up to 8th grade | 0.358 | 0.352 | 0.365 | -0.012 |
| Above 8th grade | 0.242 | 0.249 | 0.235 | 0.014 |
| *Mother's characteristics* | | | | |
| Mother's age[#] | 34.520 | 34.508 | 34.533 | -0.025 |
| No Education | 0.521 | 0.511 | 0.533 | -0.022** |
| Up to 8th Grade | 0.312 | 0.311 | 0.312 | -0.001 |
| Above 8th grade | 0.167 | 0.178 | 0.155 | 0.023*** |
| Engaged in work outside home | 0.294 | 0.288 | 0.3 | -0.013 |

*Note. The variables marked with a "#" are continuous variables and hence mean is considered instead of proportion.*

Appendix figure F1: Gender gap over years across states

(a) Andhra Pradesh
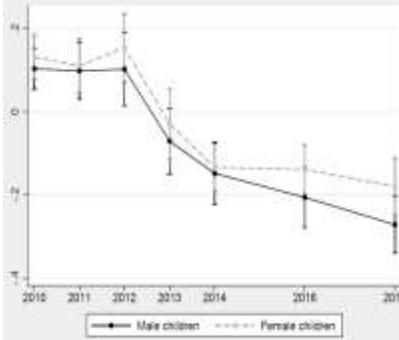

(b) Assam
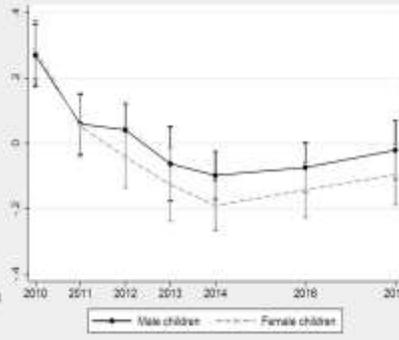

(c) Chhattisgarh
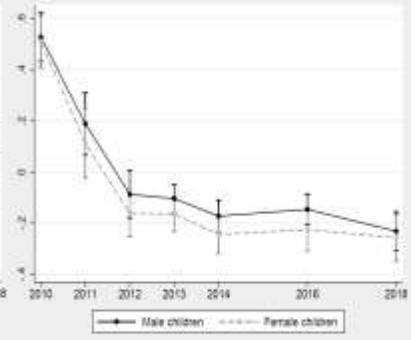

(d) Goa
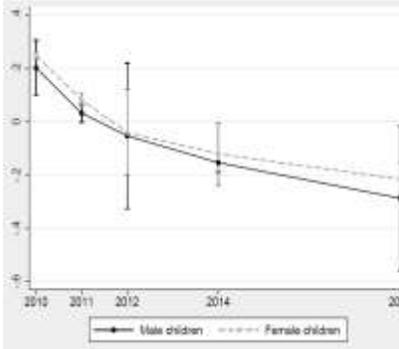

(e) Gujarat
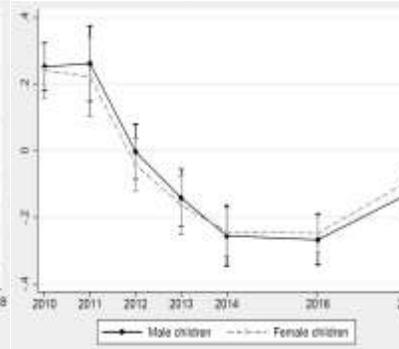

(f) Haryana
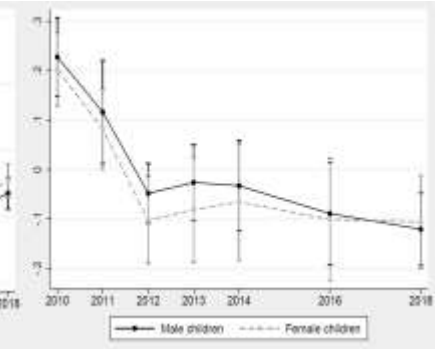

(g) Himachal Pradesh
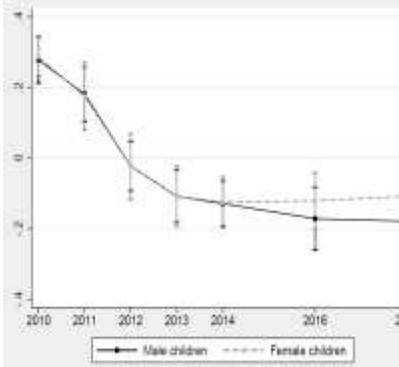

(h) Jammu and Kashmir
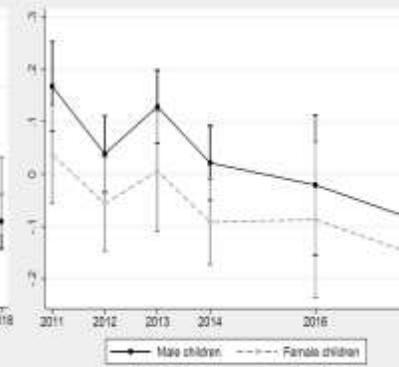

(i) Karnataka
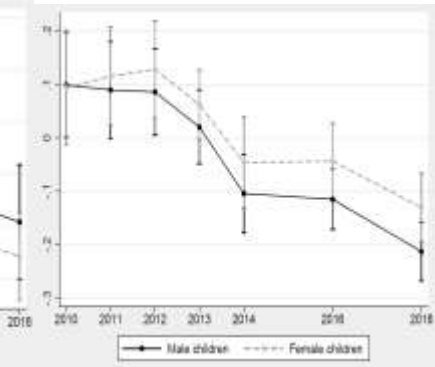

(j) Kerala 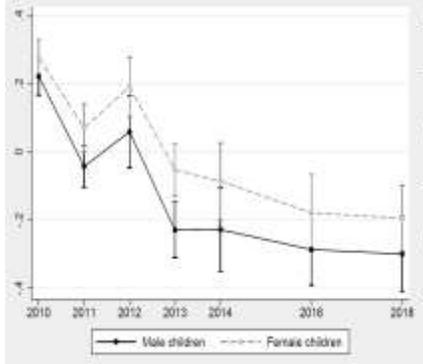 (k) Maharashtra 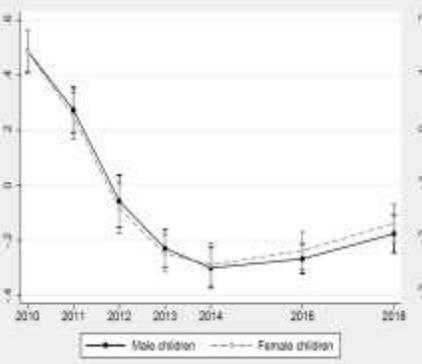 (l) North-East (excluding Assam) 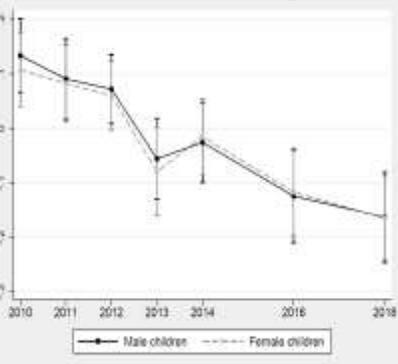

(m) Orissa 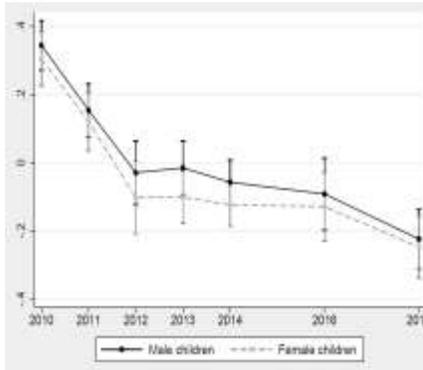 (n) Pondichéry 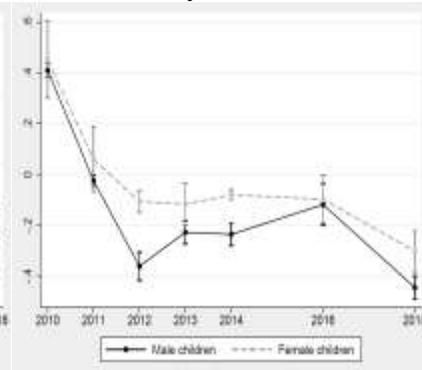 (o) Punjab 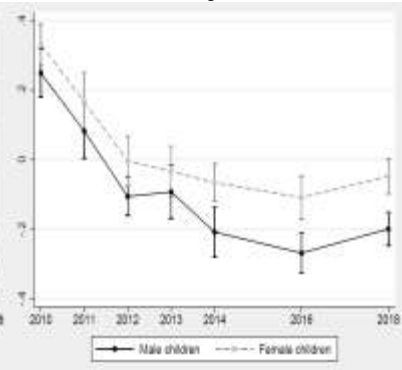

(p) Tamil Nadu 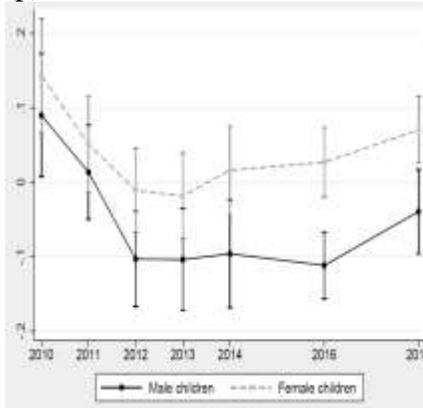 (q) Uttaranchal 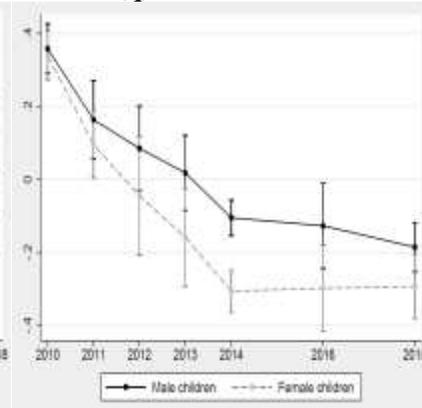 (r) West Bengal 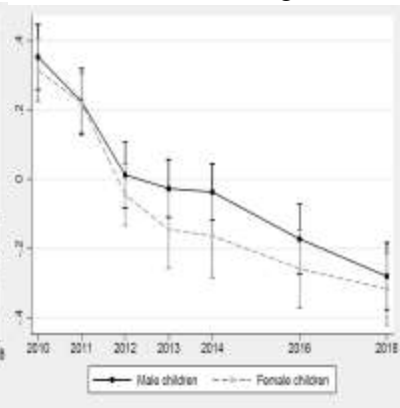

*Note: 95% Confidence intervals are calculated by clustering the standard errors at the district level are plotted along with the marginal effects. The sample includes children in the age group 8-16 years from ASER 2010 to 2018 from the respective states.*

Supplementary table ST1: Estimation of reading and mathematics scores from ASER dataset

|  | Reading scores | | | Mathematics score | | |
| --- | --- | --- | --- | --- | --- | --- |
|  | (1) | (2) | (3) | (4) | (5) | (6) |
| Female child | 0.004 | 0.003 | 0.002 | -0.078*** | -0.078*** | -0.079*** |
|  | (0.004) | (0.004) | (0.004) | (0.005) | (0.005) | (0.005) |
| Age of the child | 0.160*** | 0.160*** | 0.160*** | 0.155*** | 0.155*** | 0.157*** |
|  | (0.002) | (0.002) | (0.002) | (0.001) | (0.001) | (0.001) |
| Enrolled in school | 1.048*** | 1.046*** | 1.028*** | 0.995*** | 0.993*** | 0.974*** |
|  | (0.017) | (0.017) | (0.015) | (0.016) | (0.015) | (0.013) |
| Resides in cemented house | 0.100*** | 0.096*** | 0.112*** | 0.127*** | 0.125*** | 0.137*** |
|  | (0.007) | (0.006) | (0.004) | (0.009) | (0.008) | (0.005) |
| Has a toilet | 0.167*** | 0.164*** | 0.133*** | 0.207*** | 0.202*** | 0.151*** |
|  | (0.008) | (0.008) | (0.006) | (0.010) | (0.009) | (0.006) |
| Has a television | 0.195*** | 0.189*** | 0.157*** | 0.172*** | 0.168*** | 0.157*** |
|  | (0.007) | (0.007) | (0.005) | (0.009) | (0.009) | (0.005) |
| Uses computer | 0.160*** | 0.156*** | 0.138*** | 0.206*** | 0.202*** | 0.185*** |
|  | (0.004) | (0.004) | (0.004) | (0.005) | (0.005) | (0.005) |
| Has a mobile phone | 0.145*** | 0.143*** | 0.161*** | 0.138*** | 0.136*** | 0.144*** |
|  | (0.006) | (0.005) | (0.005) | (0.006) | (0.006) | (0.005) |
| Household size | -0.010*** | -0.010*** | -0.006*** | -0.010*** | -0.010*** | -0.006*** |
|  | (0.001) | (0.001) | (0.001) | (0.001) | (0.001) | (0.001) |
| Village has cemented road |  | 0.016** | 0.029*** |  | -0.006 | 0.027*** |
|  |  | (0.007) | (0.005) |  | (0.008) | (0.006) |
| Village has post office |  | 0.035*** | 0.024*** |  | 0.041*** | 0.034*** |
|  |  | (0.006) | (0.004) |  | (0.006) | (0.005) |
| Village has bank |  | -0.001 | -0.000 |  | 0.003 | -0.000 |
|  |  | (0.005) | (0.004) |  | (0.006) | (0.004) |
| Village has government primary health clinic |  | 0.013** | -0.006 |  | 0.017** | -0.007 |
|  |  | (0.006) | (0.004) |  | (0.007) | (0.004) |
| Village has private primary health clinic |  | -0.005 | -0.008** |  | -0.023*** | -0.008* |
|  |  | (0.005) | (0.004) |  | (0.007) | (0.004) |
| Village has private school |  | 0.005 | 0.031*** |  | 0.019** | 0.029*** |
|  |  | (0.006) | (0.005) |  | (0.007) | (0.005) |
| State fixed effects | Yes | Yes | Yes | Yes | Yes | Yes |
| Year fixed effects | Yes | Yes | Yes | Yes | Yes | Yes |
| Constant | -2.997*** | -3.023*** | -2.945*** | -2.762*** | -2.776*** | -2.594*** |
|  | (0.035) | (0.035) | (0.036) | (0.030) | (0.030) | (0.035) |
| Observations | 2,241,960 | 2,241,960 | 2,241,960 | 2,235,782 | 2,235,782 | 2,235,782 |
| R-squared | 0.264 | 0.265 | 0.275 | 0.274 | 0.275 | 0.294 |

Supplementary figure SF1: Estimations for gender gap in reading across children of same cohort

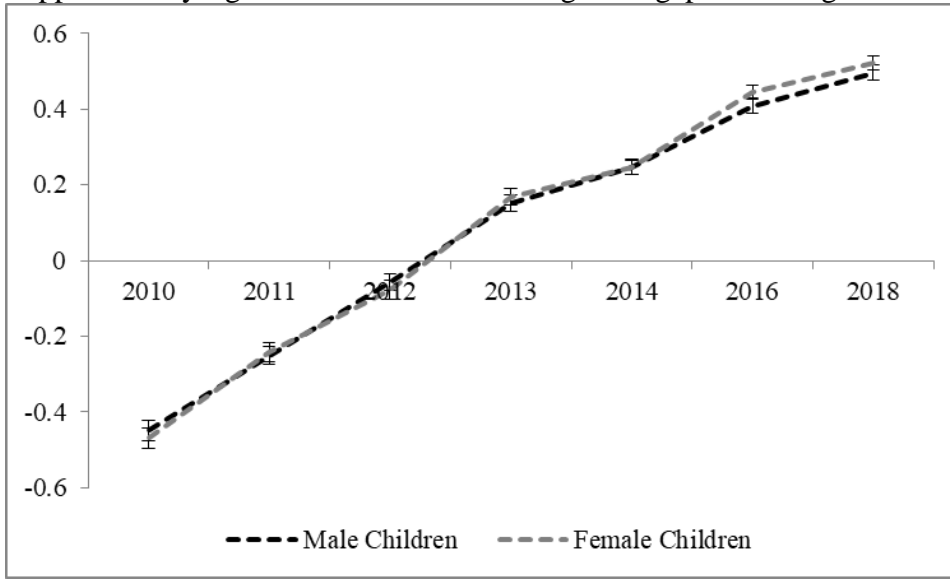

Supplementary figure SF2: Estimations for gender gap in mathematics across children of same cohort

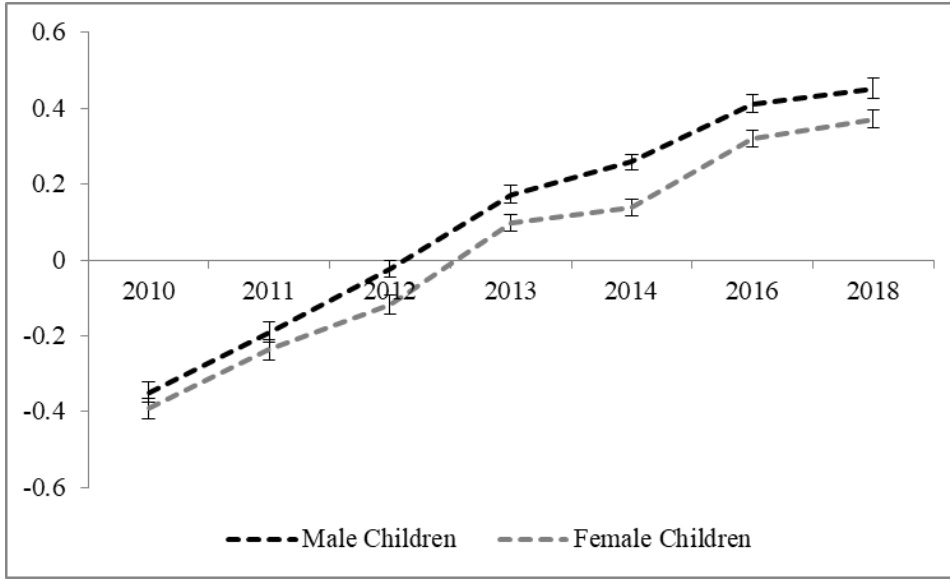